\documentstyle[epsf]{mn}

\newif\ifAMStwofonts

\def\simlt{\lower.5ex\hbox{$\; \buildrel < \over \sim \;$}}
\def\simgt{\lower.5ex\hbox{$\; \buildrel > \over \sim \;$}}

\def\lsim{ \lower .75ex \hbox{$\sim$} \llap{\raise .27ex \hbox{$<$}} }
\def\lsim{ \lower .75ex \hbox{$\sim$} \llap{\raise .27ex \hbox{$<$}} }

\def\be{\begin{equation}}
\def\ee{\end{equation}}

\def\bee{\begin{eqnarray}}
\def\eee{\end{eqnarray}}

  \ifCUPmtlplainloaded \else
  \fi
  \ifAMStwofonts
    \ifCUPmtlplainloaded \else
      \NewSymbolFont{upmath} {eurm10}
      \NewSymbolFont{AMSa} {msam10}
      \NewMathSymbol{\upi}     {0}{upmath}{19}
      \NewMathSymbol{\umu}     {0}{upmath}{16}
      \NewMathSymbol{\upartial}{0}{upmath}{40}
      \NewMathSymbol{\leqslant}{3}{AMSa}{36}
      \NewMathSymbol{\geqslant}{3}{AMSa}{3E}

    \fi
  \fi
  \addtoversion{normal}{\mathit}{cmr}{m}{it}
  \addtoversion{bold}{\mathit}{cmr}{bx}{it}

  \ifAMStwofonts
    \ifCUPmtlplainloaded \else
      \UseAMStwoboldmath
      \makeatletter
      \new@mathgroup\upmath@group
      \define@mathgroup\mv@normal\upmath@group{eur}{m}{n}
      \define@mathgroup\mv@bold\upmath@group{eur}{b}{n}
      \edef\UPM{\hexnumber\upmath@group}
      \new@mathgroup\amsa@group
      \define@mathgroup\mv@normal\amsa@group{msa}{m}{n}
      \define@mathgroup\mv@bold\amsa@group{msa}{m}{n}
      \edef\AMSa{\hexnumber\amsa@group}
      \makeatother
      \mathchardef\upi="0\UPM19
      \mathchardef\umu="0\UPM16
      \mathchardef\upartial="0\UPM40
      \mathchardef\leqslant="3\AMSa36
      \mathchardef\geqslant="3\AMSa3E

    \fi
  \fi

  \ifAMStwofonts
    \ifCUPmtlplainloaded \else
      \DeclareSymbolFont{UPM}{U}{eur}{m}{n}
      \SetSymbolFont{UPM}{bold}{U}{eur}{b}{n}
      \DeclareSymbolFont{AMSa}{U}{msa}{m}{n}
      \DeclareMathSymbol{\upi}{0}{UPM}{"19}
      \DeclareMathSymbol{\umu}{0}{UPM}{"16}
      \DeclareMathSymbol{\upartial}{0}{UPM}{"40}
      \DeclareMathSymbol{\leqslant}{3}{AMSa}{"36}
      \DeclareMathSymbol{\geqslant}{3}{AMSa}{"3E}

    \fi
  \fi
\ifCUPmtlplainloaded \else
  \ifAMStwofonts \else 
    \def\upi{\pi}
    \def\umu{\mu}
    \def\upartial{\partial}
  \fi
\fi

\title[Dynamical friction in cores and cusps]
{Does the Fornax dwarf spheroidal have a central cusp or core?}
\author[Tobias Goerdt, Ben Moore, J. I. Read, Joachim Stadel and Marcel Zemp]
{Tobias Goerdt$^1$\thanks{tgoerdt@physik.unizh.ch}, Ben Moore$^1$, J. I.
Read$^1$, Joachim Stadel$^1$ and Marcel Zemp$^{1,2}$\\
$^1$ Institute for Theoretical Physics, University of Z\"urich,
Winterthurerstrasse 190, CH-8057 Z\"urich, Switzerland\\
$^2$ Institute of Astronomy, ETH Z\"urich, ETH H\"onggerberg HPF D6, CH-8093
Z\"urich, Switzerland}
\date{Draft version \today}
\pagerange{\pageref{firstpage}--\pageref{lastpage}}
\pubyear{2006}

\begin{document}

\maketitle

\label{firstpage}

\begin{abstract}
The dark matter dominated Fornax dwarf spheroidal has five globular clusters
orbiting at $\sim 1$\,kpc from its centre. In a cuspy CDM halo the globulars
would sink to the centre from their current positions within a few Gyrs,
presenting a puzzle as to why they survive undigested at the present epoch. We
show that a solution to this timing problem is to adopt a cored dark matter
halo. We use numerical simulations and analytic calculations to show that,
under these conditions, the sinking time becomes many Hubble times; the
globulars effectively stall at the dark matter core radius. We conclude that
the Fornax dwarf spheroidal has a shallow inner density profile with a core
radius constrained by the observed positions of its globular clusters. If the
phase space density of the core is primordial then it implies a warm dark
matter particle and gives an upper limit to its mass of $\sim 0.5$\,keV,
consistent with that required to significantly alleviate the substructure
problem.

\end{abstract}




\section{Introduction}

The Fornax dwarf spheroidal is a dark matter dominated satellite orbiting the 
Milky Way. It has five globular clusters that are at a projected distance from
the centre of 1.60, 1.05, 0.43, 0.24 and 1.43\,kpc \cite{mackey} as well as
further substructure at a projected distance of 0.67\,kpc \cite{coleman}. These
star clusters move within a dense background of dark matter and should
therefore be affected by dynamical friction, causing them to lose energy and
spiral to the centre of the galaxy. We will show later that, if Fornax has a
cosmologically consistent density distribution of dark matter, the orbital
decay timescale of these objects from their current positions is
$\lsim$\,5\,Gyr. This is much shorter than the age of the host galaxy,
presenting us with the puzzle of why these five globulars have not merged
together at the centre forming a single nucleus \cite{ostriker,tremaine}. 

Several groups have studied the origin of nuclei in galaxies: e.g. Lotz et
al.\,(2001) carried out Monte Carlo simulations, which show that some, but not
all, of the nuclei of dwarf elliptical galaxies could indeed have formed
through coalescence of their globular clusters. Additionally they observed
several dE galaxies and found out that within the inner few scale lengths,
their sample appeared to be depleted of bright clusters. Oh and Lin\,(2000)
used numerical simulations to show that in dwarf galaxies with relatively weak
external tidal perturbations, dynamical friction can lead to significant
orbital decay of globular clusters and the formation of compact nuclei within a
Hubble timescale. 

Oh, Lin and Richer\,(2000) gave two possible models for the observed spatial
distribution of Fornax globulars. One possibility they proposed is that the
dark matter consists of massive black holes which transfer energy to the
globulars, preventing them from sinking to the centre of the galaxy. Another
possibility they investigated was to postulate a strong tidal interaction
between the Milky Way and Fornax which also could inject energy into their
orbits and the central core of the dSph. This latter idea is probably ruled out
due to the proper motion observations of Fornax \cite{dinescu} which suggest it
is already at closest approach on an extended orbit which never takes it close
to the Milky Way.

Here we investigate another possibility for the lack of a nucleus in Fornax,
namely that the central dark matter distribution has a very shallow cusp or
core which dramatically increases the dynamical friction sinking timescale
\cite{hernandez}. This would be inconsistent with dark haloes that form within
the CDM cosmology which have cusps steeper than -1 on all mass scales from
$10^{-6}$M$_\odot - 10^{15}$M$_\odot$ \cite{dubinski,diemand}.

Controversial evidence for cored mass distributions in dwarf spiral galaxies
has been debated for over a decade \cite{moore2}. The inner structure of
spheroidal galaxies is harder to determine, however Kleyna et al.\,(2003)
claimed that the second peak in the stellar number density in the nearby Ursa
Minor dwarf spheroidal (UMi dSph), is incompatible with cusped cold dark matter
haloes. With their observations they show that this substructure has a cold
kinematical signature and that its radial velocity with respect to its host
galaxy is very small. Such a cold configuration could only survive intact if
the stars orbited within a cored mass distribution where the orbital
frequencies are all identical (harmonic potential) and phase mixing does not
occur.

The stellar kinematical data for Fornax suggest that it is dark matter
dominated with a mass to light ratio of order 20 within its optical extent. Due
to the uncertainty on the orbital anisotropy, the mass distribution can only be
weakly constrained - the data is consistent with either cusped or cored density
distributions \cite{lokas2}. However the normalisation (or mass within the
central 1\,kpc) is better constrained. In the inner $\sim$1\,kpc
of a cored halo the mean density is approximately six times lower than in a
cusped halo. Furthermore the velocity distribution function of the background
particles is hotter than a cusped halo. These facts conspire to significantly
increase the dynamical friction timescale in a cored mass distribution.

In this paper we construct cored and cuspy dark matter potentials and calculate
orbital decay and sinking times using high resolution numerical simulations
together with analytic calculations ~\cite{chandrasekhar}. The haloes are
consistent with the kinematical data for Fornax. We follow circular and
eccentric orbits of single and multiple globular clusters. Although many
dynamical friction studies have been carried out before
\cite{white,hernquist,capuzzo}, we are not aware of any studies within 
constant density cores at the resolution used in this paper, although the
recent study explored the effects of sinking objects on various cusp structures
\cite{merritt}. In section 2 we present the numerical methods we used, the
analytical computation of the sinking times which are compared to our high
resolution numerical simulations. In section 3 we discuss our results and draw
our conclusions.

\section{Results}

We carry out a series of self consistent simulations to examine the orbital
behaviour of massive particles moving within a dark matter or stellar
background. We use the parallel multi-stepping N-body tree-code, pkdgrav2,
developed by Joachim Stadel \cite{stadel}. We construct stable particle haloes
using the techniques developed by Kazantzidis, Magorrian \& Moore (2004). These
models have density distributions that are described by the $\alpha, \beta,
\gamma$ law \cite{abc}:
\begin{equation}
\rho(r)=\frac{\rho_0} {\left({\frac{r}{r_{\rm s}}}\right)^\gamma \left[{1 +
\left({\frac{r}{r_{\rm s}}}\right)^{\alpha}}\right]^{\frac{\beta - \gamma}
{\alpha}}}
\label{eq:cusp}
\end{equation}

For our simulations we used 
``NFW-like'' haloes \cite{nfw,moore} with $\alpha = 0.5-1.5$, $\beta = 3.0$ and
$\gamma = 0.5-1.5$, or cored haloes with $\alpha = 0.5-1.5$, $\beta = 3.0$ and
$\gamma = 0.0$. In the former case we have $\rho_0 = 0.0058$\,M$_\odot$/pc$^3$
and $r_{\rm s}= 2.4$\,kpc. This cuspy halo has a virial mass of $2.0 \times
10^9$\,M$_{\odot}$. The concentration parameter is 15 but our results would not
change with a lower concentration, since in either case we are within the
asymptotic cusp part of the density profile. 

We use a three shell model \cite{zemp}; $10^5$ particles for the innermost
sphere with 100 pc radius, $10^5$ particles for the shell between 100 and
500\,pc and $10^5$ particles for the rest of the halo. The softening lengths of
the particles in these shells are 1, 10 and 100\,pc respectively. The results
were found not to be sensitive to these values. The particle masses are
58\,M$_\odot$, 569\,M$_\odot$ and 3.2 $\times 10^4$\,M$_\odot$. These models
are stable in isolation but allow us to achieve very high resolution at the
halo centre where we wish to follow the dynamical friction.

\begin{figure}
\begin{center}
\epsfxsize=8.5cm
\epsfysize=7.08cm
\epsffile{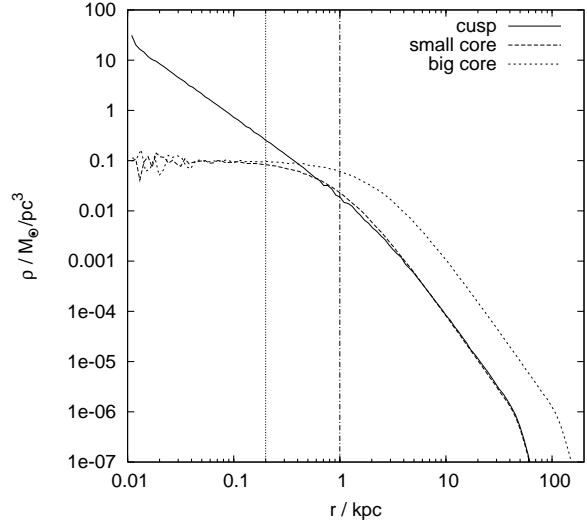}
\end{center}
\caption{The initial radial density profiles for the three different haloes
used in the simulations. The vertical lines indicate the size of the core of
the cored haloes.}
\label{figrdp}
\end{figure}

For a small cored halo we have $\rho_0 = 0.10$\,M$_\odot$/pc$^3$ and
$r_{\rm s} = 0.91$\,kpc (n.b. the radius at which the slope of the density
profile is shallower than -0.1 is approximately 200\,pc which defines the
constant density region in this model). This halo has a virial mass of
$2.0 \times 10^9$\,M$_\odot$ and the concentration parameter is 40. Again, we
use a three shell model that has $10^5$ particles for the innermost sphere with
300\,pc radius, $10^5$ particles for the shell between 0.3 and 1.1\,kpc and $3
\times 10^5$ particles for the rest of the halo. The softening lengths of the
particles in these shells are 3, 30 and 300\,pc respectively. The particle
masses are 89\,M$_\odot$, 1640\,M$_\odot$ and 7572\,M$_\odot$. For a big cored
dark matter halo we have basically the same parameters as for the halo with the
small core, except for the scale length $r_s = 2.2$\,kpc (here the constant
density region is approximately 1\,kpc), the virial mass $M_{\rm vir} = 3.0
\times 10^{10}$\,M$_\odot$ and the particle masses, which are in this case
106\,M$_\odot$, 3625\,M$_\odot$ and $1.2 \times 10^5$\,M$_{\odot}$. The density
profiles of these three haloes are shown in figure \ref{figrdp}. 

The density profiles agree fairly well with the constraints made by
observations of Fornax \cite{lokas2,walker}. We did actually perform the same
simulations and treatment as described in the following with the very haloes
proposed by Lokas (2002) with the same results. We also repeated several of our
simulations with haloes modelled with ten times as many particles than
described above. These high resolution runs show exactly the same features as
the low resolution runs, but with less noise.

Where available, we will present the high resolution graphs in this paper. The
globular clusters are modelled as single particles of mass $M_{\rm GC} = 2
\times 10^5$\,M$_{\odot}$ with a softening of 10\,pc. We do not expect our
conclusions to change if we used a particle model for each globular since they
are stable against tidal disruption within Fornax. We start the globulars
outside the core, mostly on circular orbits and let them orbit, expecting them
to spiral in to the centre of their respective host haloes due to dynamical
friction. The distance from the centre of the host halo as a function of time,
$r(t)$, can be computed using Chandrasekhar's dynamical friction formula
\cite{binney}, which is given by:
\begin{eqnarray}
F &=& -{4 \pi {\rm ln} \Lambda (r) \rho(r) G M^2_{\rm GC} \over v^2_{\rm c}(r)}
\nonumber \\ & & \left[{{\rm erf} \left({{v_{\rm c}(r) \over \sqrt{2}
\sigma(r)}}\right) - {2 v_{\rm c}(r)\over \sqrt{2 \pi} \sigma(r)} e^{-{v_{\rm
c}^2(r) \over 2 \sigma^2(r)}} }\right],
\label{eq:dff}
\end{eqnarray}
which gives the force acting on the massive particle crossing the halo. 
The density profile $\rho(r)$ is given by our equation \ref{eq:cusp}, and we
assume that the velocity distribution is isotropic and Maxwellian at all radii.
Of course this assumption does not hold, but is good enough for our purposes
\cite{stelios}. We can then easily calculate the velocity dispersion using the
Jeans equation:
\begin{equation}
\sigma^2(r) = {1 \over \rho(r)}\int^{\infty}_{r}{{M(r') \rho(r')\over r'^2}
dr'}.
\end{equation}

\begin{figure}
\begin{center}
\epsfxsize=8.5cm
\epsfysize=7.08cm
\epsffile{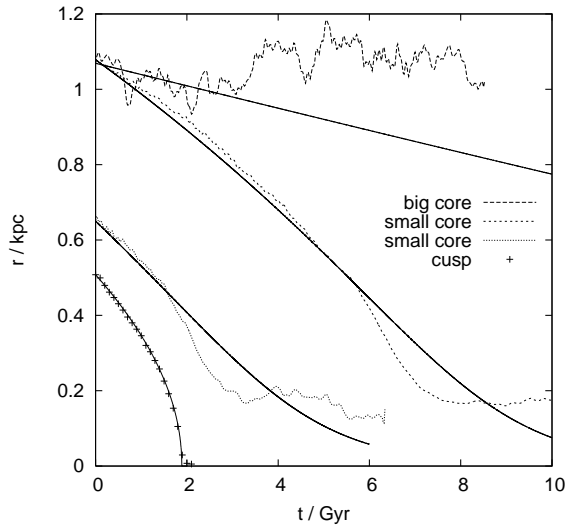}
\end{center}
\caption{Radial distance of the single globular cluster from the centre of its
host halo as a function of time. We start the calculations with the globular at
different initial radii for clarity. Solid curves are the analytic estimates,
dashed curves are from the numerical simulations.}
\label{figcpco}
\end{figure}

We find similar sinking times for eccentric orbits, therefore for brevity we
show only the circular orbits in this paper and leave the detailed parameter
space study for a future paper which explores the technical aspects of
dynamical friction in structures with different density profiles. Additionally
we assume, that $M_{\rm GC} \gg$ $m_{\rm par}$. This is a little problematic in
the case of the cuspy and the big cored potential because the particles in the
outermost shell have $m_{\rm par} = 3 \times 10^4$\,M$_\odot$ and $m_{\rm par}
= 1.2 \times 10^5$\,M$_\odot$, respectively. However, these particles rarely
penetrate the innermost 0.5\,kpc of the halo. In equation \ref{eq:dff}, ${\rm
ln} \Lambda (r)$, is the Coulomb logarithm:
\begin{equation}
{\rm ln} \Lambda (r) = {b_{\rm max} \sigma^2(r) \over GM_{\rm GC}},
\end{equation}
in this definition $b_{\rm max}$ is the largest impact parameter to be
considered. This parameter is defined by one of the assumptions
Chandrasekhar made while deriving the above dynamical friction formula: The
intruder must be moving through a medium with constant density, therefore
$b_{\rm max}$ is the greatest distance for which this is still valid. We keep
$b_{\rm max}$ as a free parameter when fitting our analytic formulae to the
simulations. We find for the cuspy haloes $b_{\rm max} = 0.25$\,kpc and for the
cored ones $b_{\rm max} = 1.0$\,kpc. In this equation  $v_{\rm c}(r)$ is the
circular velocity at radius $r$. The force exerted by dynamical friction on the
perturber is tangential with respect to its movement and thus causes the
cluster to lose angular momentum per unit mass at a rate
\begin{eqnarray}
{dL \over dt} & = & {F r \over M_{\rm GC}} = -{4 \pi {\rm ln} \Lambda (r)
\rho(r) G M_{\rm GC} r \over v^2_{\rm c}(r)} \nonumber \\ & & \left[{{\rm erf
}\left({v_{\rm c}(r) \over \sqrt{2}\sigma(r)}\right) - {2 v_{\rm c}(r) \over
\sqrt{2 \pi} \sigma(r)} e^{-{v_{\rm c}^2(r) \over 2 \sigma^2(r)}}}\right].
\label{eq:L1}
\end{eqnarray}
Since the cluster continues to orbit at a speed $v_{\rm c}(r)$ as it spirals to
the centre, its angular momentum per unit mass at radius $r$ is at all times
$L = rv_{\rm c}(r)$. Substituting the time derivative of this expression into
equation \ref{eq:L1} we obtain
\begin{eqnarray}
{dr \over dt} & = & -{4 \pi {\rm ln} \Lambda (r) \rho(r) G M_{\rm GC} r \over
v^2_{\rm c}(r) {d(rv_{\rm c}(r)) \over dr}} \nonumber \\ & & \left[{{\rm
erf}\left({v_{\rm c}(r) \over \sqrt{2} \sigma(r)}\right) - {2 v_{\rm c}(r)
\over \sqrt{2 \pi} \sigma(r)} e^{-{v_{\rm c}^2(r) \over 2
\sigma^2(r)}}}\right].
\label{eq:L2}
\end{eqnarray}
Substituting values for the initial radii we obtain the analytical curves drawn
in figure \ref{figcpco} plotted on top of the results from the numerical
simulations.

For the cuspy potential the analytic calculation agrees very well with the
numerical simulation. Haloes with a core give a poorer agreement. After an
initial sinking rate that agrees well with the analytic expectation, the
globulars sink faster as they approach twice the core radius, and then stop
sinking at the core radius. The analytic formula predicts a continued, but slow
infall to the centre. This resonance/scattering effect will be investigated in
a more detailed paper \cite{goerdt}. We note, however, that it is not trivially
due to the fact that the globular is of comparable mass to that enclosed by its
orbit; the radius at which $M(r)=M_{GC}$ is $\sim$ three times smaller than the
core radius (see Figure \ref{figcore}). The stalling results are apparent in
both of the cored halo simulations (small core and big core). We conclude that
the presence of a central density core leads to the infall of the clusters
stopping at the core radius; the problem is in this sense scalable. 

\subsection{Particle noise and halo centring}
It is difficult to define the centre of the constant density core. The
potential has a negative minimum, but $\sqrt{N}$ noise perturbations can have
deeper potentials than the core itself. Thus defining the centre using the most
bound particle gives a large error in determining the centre. We found that a
shrinking spheres method can work, so long as one takes the centre of mass of a
sphere containing most of the core. If you continue to shrink the sphere based
on a small number ($<10^4$) particles, then this also picks out the largest
Poisson fluctuation in the core. The centre defined using $\sim 10^5$ particles
gives a robust estimate.

In our standard resolution simulations we use $10^5$ particles in the high
resolution region each of mass 89\,M$_{\odot}$. Simulating the entire halo at
this resolution would require $\sim 4 \times 10^7$  particles. However, as
discussed above, $\sqrt{N}$ noise can be substantial, even at this resolution.
This may introduce spurious heating, preventing the globulars from sinking. We
can investigate this by examining the orbit of the globular cluster. For a
perfectly smooth spherical potential the orbit of the globular cluster would
always remain in its initial orbital plane. Fluctuations from particle
representation will cause deviations from this plane. Once the globular reaches
the high density centre in the cases of the low resolution runs fluctuations in
the orbital plane become very large (fluctuations in L$_z$/L $\sim $0.5$\%$) -
the relaxation time is very short and acts to counter dynamical friction. We
therefore carried out simulations with ten times as many particles. At this
resolution the fluctuations are greatly reduced  (fluctuations in L$_z$/L $\sim
$0.05$\%$).

\subsection{Multiple globular clusters}
Finally we re-ran these simulations using five globulars to study the effect
of having multiple sinking objects. Perhaps interactions between the globulars
themselves may prevent them from sinking to the central cusp and merging.
We distribute the globulars randomly, what position and plane of the orbit
concerns, with distances to the centre between 0.2 and 0.8\,kpc. The clusters
are again placed on circular orbits around the centre of their host halo and
are evolved with pkdgrav2.

\begin{figure}
\begin{center}
\epsfxsize=8.5cm
\epsfysize=7.08cm
\epsffile{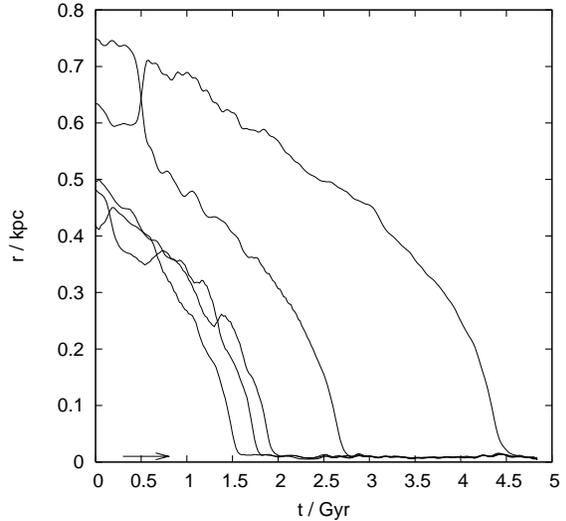}
\end{center}
\caption{Radial distance of the five globular clusters from the centre of their
host halo as function of time, as they orbit within a cusped density
distribution. The arrow indicates the radius at which $M_{\rm GC} = M(r)$.}
\label{figcusp}
\end{figure}

\begin{figure}
\begin{center}
\epsfxsize=8.5cm
\epsfysize=7.08cm
\epsffile{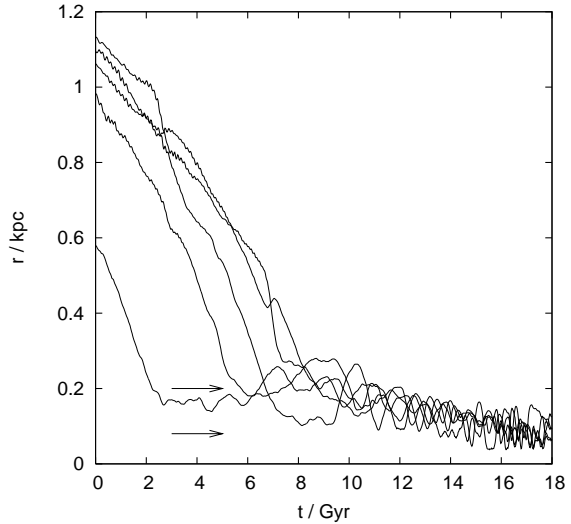}
\end{center}
\caption{Radial distance of the five globular clusters from the centre of the
host halo as function of time within the cored potential. The upper arrow
indicates the size of the core and the lower arrow indicates the radius at
which $M_{\rm GC} = M(r)$.}
\label{figcore}
\end{figure}

Interestingly, the clusters do not prevent one another from falling to the
centre, but instead create an interesting prediction. Figure \ref{figcusp}
shows the infall as a function of time for the five globulars in the cuspy dark
matter halo. Notice that all of the clusters fall to the centre. However,
clusters which start out at very similar radii arrive $\sim 1$Gyr apart. At any
given time, even for very similar initial conditions, the clusters occupy a
range of radii. By contrast, for the case with a central dark matter core (see
figure \ref{figcore}), although the globulars still arrive at different times
due to interactions, they stall at the core radius; they do not sink to the
centre even within 20\,Gyrs\footnote{Recall that this stalling behaviour was
also present in the run with a larger core size and so is not peculiar to the
small core density profile.}. Thus, if Fornax does have a central constant
density core we should expect the clusters to stall at some minimum radius. No
globular cluster could possibly get any closer to the centre of the halo
than the core radius. The lower limit of the core size is constrained by the
smallest observed projected cluster distances to be 0.24\,kpc.

\begin{figure}
\begin{center}
\epsfxsize=8.5cm
\epsfysize=7.08cm
\epsffile{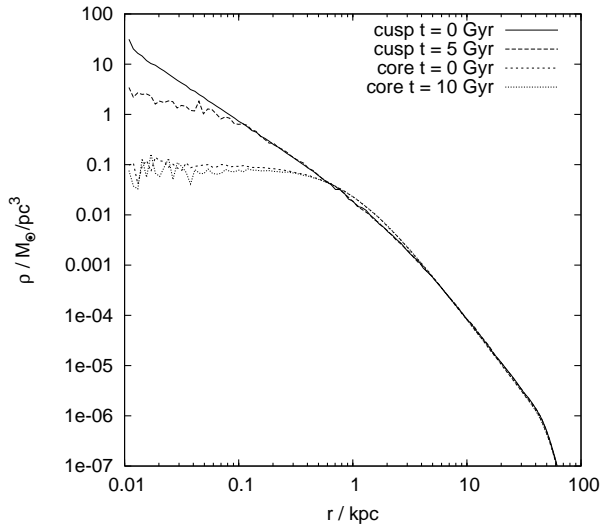}
\end{center}
\caption{Radial density profile of the dark matter haloes. Only the dark
matter is shown, the mass of the globular clusters is neglected. The initial
conditions are compared with the final state of the system. An isolated halo,
evolved for the same time, has exactly the same density profile as the initial
conditions.}
\label{figrep}
\end{figure}

Finally, one can see in figure \ref{figrep} that for the case where there is
a cusp and the globular clusters do spiral in, they displace the dark matter
from the centre. Dark matter particles move out of the nucleus. The density
of the dark matter in the centre drops by more than an order of magnitude, an 
effect that does not happen in cored haloes. 
We note that several authors have recently studied this process in detail.
For example \cite{amr}\cite{merritt} study the change in an initial cuspy 
density profile due to the frictional effects of sinking objects.
This process allows initially cusped density profiles to be
transformed into nearly harmonic cores as the globulars fall in. For the case
of the five globulars in the Fornax halo, the maximum initial central density
slope for which this can occur is approximately 0.5. Thus in order for these
data to be consistent with a CDM halo, the central cusp must have been modified
and flattened to a slope shallower than this value. Exotic scenarios which
might achieve this include infall and subsequent blowout of massive gas clouds
or star clusters \cite{read}.

\section{Conclusions}

The Fornax dwarf spheroidal has five globular clusters orbiting at a projected
radius  of $\sim$ 1\,kpc from its centre. Using a cuspy CDM potential with
central slope steeper than 0.5 and normalised to match the inferred properties
from the kinematical data, we find that these globulars would all sink to the
halo centre within 5 Gyrs.

By contrast, we showed that if Fornax has a constant density central core then
the dynamical friction time becomes arbitrarily long - the globulars stop
sinking at the edge of the core, thus the present position of the {\it
innermost} globular gives a lower limit of the core radius of the dark matter
distribution. Since CDM uniquely predicts that all haloes are cusped, this
suggests that the dark matter distribution within Fornax is not cold, but may
be warm dark matter or some other candidate. Alternatively, the mass
distribution has been modified through some exotic dynamical phenomenon such as
rapid mass loss \cite{read}. If the phase space density of the core, measured
as $Q\sim 10^{-5}$M$_\odot$\,pc$^{-3}$(kms$^{-1}$)$^{-3}$ in our model, is
primordial then it implies a warm dark matter particle of mass $\sim 0.5$\,keV
\cite{hogan,dalcanton}. This is consistent with that required to largely solve
the substructure problem \cite{moore3}.

Although the dwarf spheroidals around the Milky Way do not contain prominent
nuclei, about 30 percent of dwarf spheroidals (dE's) in clusters are nucleated.
If these nuclei form by the merging of star clusters then we must conclude that
these galaxies have cuspy mass distributions. This could be due to the fact
that transformation to dE via galaxy harassment gives an exponential
distribution of stars which usually dominate the central mass distribution.
Therefore globulars could sink via friction against the stellar background.
Most nucleated dE's are near the cluster centre where harassment is important
which supports this idea.

Our numerical simulations show that the standard Chandreskhar estimate of 
sinking timescales works well for cuspy cores but fails completely for 
cored mass distributions. In addition we showed that over $10^6$ particles
are required within the core region to suppress heating from particle noise
\cite{weinberg}. This is particularly important in a cored mass distribution
where the potential minimum is quite shallow. The fact that cored mass
distributions do not give rise to dynamical friction is an important result. We
believe that this is due to orbit-resonant scattering. This will be the subject
of a forthcoming paper \cite{goerdt}.

We have shown that a natural solution to the timing problems for Fornax's
globular clusters is a central dark matter core. A prediction of this model is
that the clusters have a well-definied minimum radius. Fornax is not alone in
showing indirect evidence for such a core. The UMi dSph galaxy also has
substructure which appears to have survived longer than is possible within a
cuspy halo. Understanding the origin of these constant density cores could be
one of the most exciting challenges facing astronomy in the next few years. 

\section*{Acknowledgements}
It is a pleasure to thank Tobias Huber for the fruitful discussions. 
Special thanks to Doug Potter for bringing to life the zBox2 supercomputer
(http://www.zbox2.org) at the University of Z\"urich, on which all of the
computations were performed.
\label{lastpage}

\end{document}